\documentclass[%
 preprint,
showpacs,preprintnumbers,
 amsmath,
 amssymb,
 aps,
 prb,
]{revtex4-1}

\usepackage{graphicx}
\usepackage{dcolumn}
\usepackage{bm}

\begin{document}

\title{Unidirectional wave propagation in media with complex principal axes}

\author{S. A. R. Horsley}
\affiliation{Department of Physics and Astronomy, University of Exeter, Stocker Road, Exeter EX4 4QL}

%
%
\begin{abstract}
In an anisotropic medium, the refractive index depends on the direction of propagation.  Zero index in a fixed direction implies a stretching of the wave to uniformity along that axis, reducing the effective number of dimensions by one.  Here we investigate two dimensional gyrotropic media where the refractive index is zero in a \emph{complex} valued direction, finding that the wave becomes an analytic function of a single complex variable $z$.  For simply connected media this analyticity implies unidirectional propagation of electromagnetic waves, similar to the edge states that occur in photonic `topological insulators'.  For a medium containing holes the propagation is no longer unidirectional.  We illustrate the sensitivity of the field to the topology of the space using an exactly solvable example.  To conclude we provide a generalization of transformation optics where a complex coordinate transformations can be used to relate ordinary anisotropic media to the recently highlighted gyrotropic ones supporting one--way edge states.
\end{abstract}
\pacs{42.25.-p, 42.70.-a, 03.50.De, 02.40.Tt}

\maketitle
%
%
There are many situations where it might be useful to force a wave to propagate in only one direction; from the transfer of power down a cable, to the communication between two antennas.  But this is difficult.  We might launch a wave rightwards down a channel, intending it to transport all of its energy to the far end.  But any obstacle in the channel inevitably generates reflection, returning part of the wave to its sender.  This less than perfect transmission occurs because almost all of the wave equations of physics possess a left--going solution for every right--going one (implied by various reciprocity theorems~\cite{potton2004}), and an obstacle provides a coupling between them.
\par
True unidirectional propagation is possible, but requires a violation of reciprocity.  There has been some fascinating work to achieve this for waves trapped at interfaces between periodic media, in both electromagnetic~\cite{haldane2008,wang2008,rechtsman2013,soljacic2014,jacobs2015} and acoustic~\cite{yang2015,mousavi2015,fleury2016} systems.  The two periodic media must share a common band gap, and a sufficient condition for unidirectional waves (so--called `edge states') to be trapped at the interface can be written in terms of a topological quantity called the Chern number~\cite{nakahara2003}, which is computed as an integral of the Berry curvature~\cite{berry1984} associated with the Bloch functions.  This recipe has its origins in the theory of the quantum Hall effect, where the Chern number appears directly in the formula for the electrical conductivity~\cite{thouless1994}, and the Berry curvature in reciprocal space can be understood as originating from the points in the Brillouin zone where the edge states join the bulk modes~\cite{hatsugai1993}.  While the theory is well developed in the condensed matter setting, in acoustics and electromagnetism it remains---to a degree---mysterious.
\par
For instance, there exist inhomogeneous \emph{aperiodic} materials in which Maxwell's equations take the form of the time independent Schr\"odinger equation for a charged particle in a uniform magnetic field~\cite{liu2015}.  Despite the formal similarity with the quantum Hall effect, it is not obvious how to adapt the known topological results~\cite{thouless1994,hatsugai1993} to this electromagnetic case.  In addition, there exist \emph{homogeneous} gyrotropic materials that support unidirectional states at an interface with a perfect conductor~\cite{davoyan2013}.  In these uniform media the Brillouin zone can be chosen arbitrarily, and again it becomes unclear if the unidirectional states can be predicted through computing anything like the Chern number.  Nevertheless, some intriguing recent work has attempted to set up a formalism for computing the Chern number in these homogeneous media~\cite{silveirinha2015,silveirinha2016}.
\par
This paper was partly inspired by the aforementioned work of Davoyan, Engheta and Silveirinha~\cite{davoyan2013,silveirinha2015,silveirinha2016}.  We focus on understanding unidirectional propagation in the same kind of homogeneous gyrotropic media as them, but look mainly at the range of parameters where the material is on the tipping point between opacity and transparency (where the unidirectional edge states occur alongside the bulk modes).  Rather than work in terms of the Chern number (as was done in~\cite{silveirinha2015,silveirinha2016}) we look for a different way to understand unidirectional propagation.  We find that when the refractive index is zero along an axis that points in a complex direction, the field becomes a function of a single complex variable $z$.
\par
This dependence on $z$ can be understood as the origin of the unidirectional propagation.  Something similar is already known to occur in the quantum Hall effect, where the ground state wave--function of a charged particle (charge $e$) in a homogeneous magnetic field $B\hat{\boldsymbol{z}}$, subject to periodic boundary conditions, is given by (see e.g.~\cite{fremling2013})
\begin{equation}
	\psi(x,y)={\rm e}^{-\pi\left(\frac{x}{a}\right)^2}\theta_{3}\left(-\frac{{\rm i}\pi z}{a},{\rm e}^{-\pi}\right)
\end{equation}
where $\theta_{3}$ is a theta function as defined in~\cite{dlmf}, and the wave is periodic in $x$ and $y$ with period $a=\sqrt{h/eB}$.  The above dependence of the wave--function on the complex variable $z=x+{\rm i}y$ is a beautiful expression of unidirectional propagation: given that the phase of the theta function determines the propagation of the wave, and admits a Taylor expansion $\sum_{n\geq0}c_{n}z^{n}=\sum_{n\geq0}c_{n}r^{n}\exp({\rm i}n\theta)$, we can see that the wave contains a series of terms, each of which has a phase that winds the same way around the origin.
\par
The main findings of this paper are; (i) that in gyrotropic media the electromagnetic field can behave as an analytic function of a complex position variable, thus exhibiting unidirectional propagation and an extreme sensitivity to the topology of the space; and (ii) that media exhibiting such unidirectional propagation can be connected with the use of complex spatial coordinates in transformation optics~\cite{horsley2015,longhi2016,horsley2016,castaldi2013,pendry1996,pendry2006,leonhardt2010}.  These results provide an alternative, arguably more intuitive method for designing homogeneous media supporting one--way electromagnetic states, and illustrate how many of the unusual features of wave propagation in these media can be understood in terms of the basic properties of analytic functions of a single complex variable.  It is possible that the theory presented in this paper may also be extended to design inhomogeneous media that support such states.
%
%
\section{Wave propagation in gyrotropic media}
\par
We shall first review the distinction between wave propagation in gyrotropic and 'ordinary' materials.  We consider an electromagnetic wave propagating in the $x$--$y$ plane, in a non--magnetic ($\mu=1$) material characterized by the following Hermitian permittivity tensor
\begin{equation}
	\boldsymbol{\epsilon}=\left(\begin{matrix}\epsilon_{1}&|a|{\rm e}^{{\rm i}\phi}&0\\|a|{\rm e}^{-{\rm i}\phi}&\epsilon_{2}&0\\0&0&\epsilon_{3}\end{matrix}\right)=\left(\begin{matrix}\boldsymbol{\epsilon}_{\parallel}&\boldsymbol{0}\\\boldsymbol{0}&\epsilon_{3}\end{matrix}\right)\label{eq:gyrotropic-medium}
\end{equation}
The term `gyrotropic' is used to indicate that this tensor is complex Hermitian rather than real symmetric, so the medium becomes `ordinary' when $\phi=0,\pi$.
\par
In such planar media the polarization separates into two simple types.  Either the magnetic field ($\boldsymbol{H}=H\hat{\boldsymbol{z}}$) or the electric field ($\boldsymbol{E}=E\hat{\boldsymbol{z}}$) point out of the plane of propagation.  In this work we consider only waves where the magnetic field points out of the plane ($H$ polarized).  Then the components of the $2\times 2$ tensor $\boldsymbol{\epsilon}_{\parallel}$ determine the propagation.  After a few manipulations of Maxwell's equations, we find that the magnetic field is subject to a modified version of the Helmholtz equation
\begin{align}
	\boldsymbol{\nabla}\cdot\boldsymbol{\epsilon}_{s}\cdot\boldsymbol{\nabla}H+k_0^{2}{\rm det}[\boldsymbol{\epsilon}_{\parallel}]H&=\Lambda_{1}\left(\hat{\boldsymbol{n}}_{1}\cdot\boldsymbol{\nabla}\right)^{2}H+\Lambda_{2}\left(\hat{\boldsymbol{n}}_{2}\cdot\boldsymbol{\nabla}\right)^{2}H+k_0^{2}\lambda_{1}\lambda_{2}H\nonumber\\
    &=0\label{eq:TMwveqn}
\end{align}
where
\[
	\boldsymbol{\epsilon}_{s}=\frac{1}{2}[\boldsymbol{\epsilon}_{\parallel}+\boldsymbol{\epsilon}_{\parallel}^{\rm T}]
\]
is the symmetric part of the tensor $\boldsymbol{\epsilon}_{\parallel}$.  Because there is a distinction between $\boldsymbol{\epsilon}_{s}$ and $\boldsymbol{\epsilon}_{\parallel}$, we have two sets of eigenvalues and principal axes to consider: $\Lambda_{i}$ and $\hat{\boldsymbol{n}}_{i}$ are the eigenvalues and principal axis directions of $\boldsymbol{\epsilon}_{s}$, while $\lambda_{i}$ are the eigenvalues of the Hermitian tensor $\boldsymbol{\epsilon}_{\parallel}$ (we define the principal axes of this tensor in the next section).
%
%
\begin{figure}[ht!]
	\includegraphics[width=16cm]{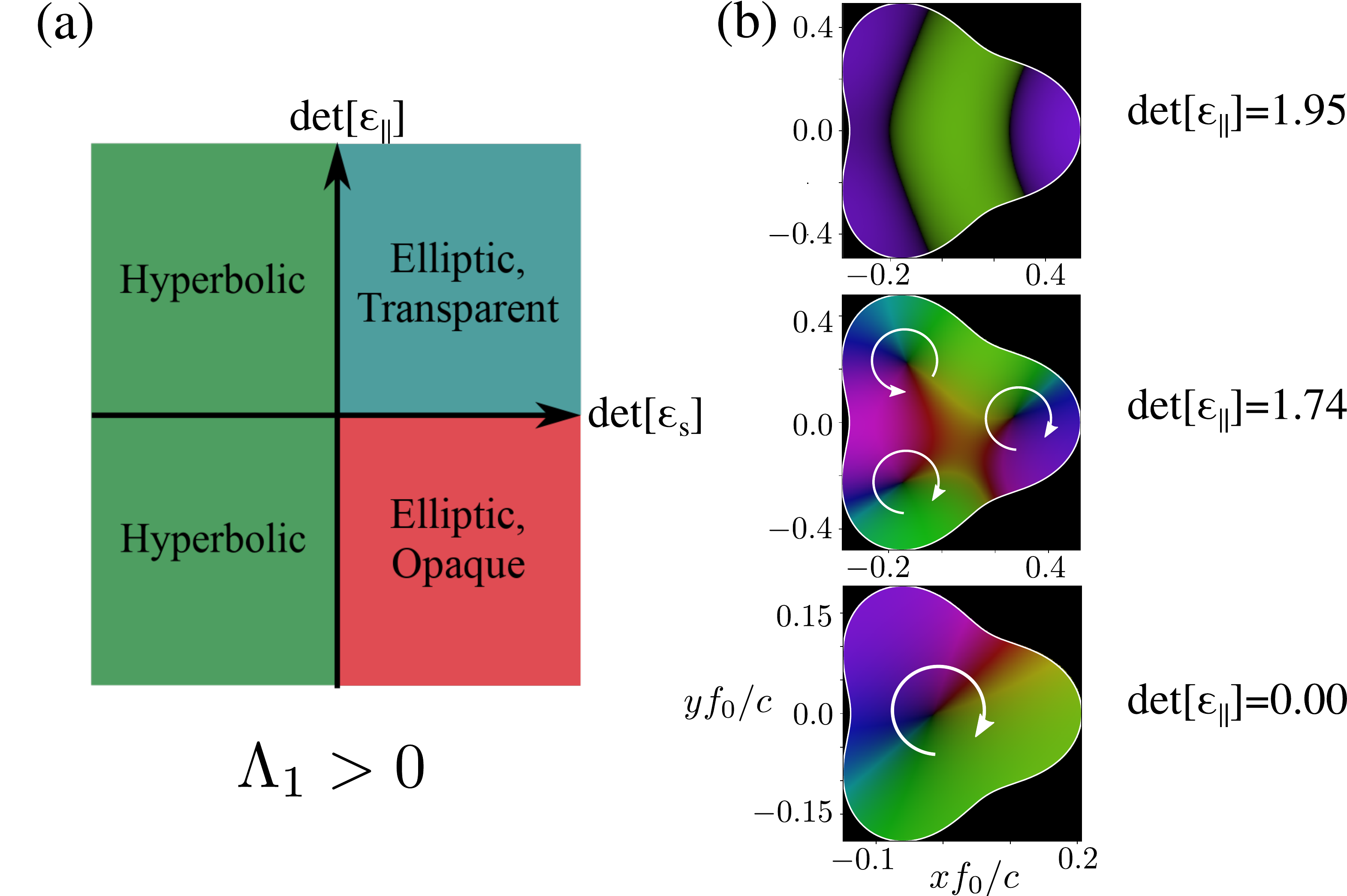}
	\caption{(a) In gyrotropic media the Helmholtz equation depends separately on the eigenvalues of $\boldsymbol{\epsilon}_{\parallel}$ and the symmetric part of this tensor $\boldsymbol{\epsilon}_{s}$, which are independent.  The plot shows the different possible properties of the media, for a fixed sign of one of the eigenvalues of $\boldsymbol{\epsilon}_{s}$.  (b) The lowest frequency ($f=f_0$) $H$ polarized eigenmode of a clover leaf shaped cavity (calculated using COMSOL multiphysics).  The three plots show the $H$ field as the determinant of $\boldsymbol{\epsilon}_{\parallel}$ is reduced to zero (arrows indicating propagation direction), for the particular parameters $\epsilon_{1}=1.5$ $\epsilon_{2}=1.3$, $\phi=0.2\pi$ and $|a|=\sqrt{\epsilon_{1}\epsilon_{2}}\times\{0.00001,0.3,1.0\}$.  Color indicates phase (advancing red, green, blue) and brightness indicates magnitude.  When $\text{det}[\boldsymbol{\epsilon}_{\parallel}]\sim0$, the field becomes a function of a single complex variable $z$, in this case with a zero at the centre of the cavity.\label{fig:gyrotropy}}
\end{figure}
\par
In `ordinary' media where the permittivity tensor (\ref{eq:gyrotropic-medium}) is real and symmetric, the eigenvalues of $\boldsymbol{\epsilon}_{s}$ and $\boldsymbol{\epsilon}_{\parallel}$ are identical $\Lambda_{i}=\lambda_{i}$.  In this case, a zero for one of these two eigenvalues reduces (\ref{eq:TMwveqn}) to the one dimensional Laplace equation (e.g. $\left(\hat{\boldsymbol{n}}_{2}\cdot\boldsymbol{\nabla}\right)^{2}H=0$), implying that in an infinite medium the magnetic field $H$ depends on only one coordinate, and is therefore uniform along one axis.  Meanwhile, in a gyrotropic medium ($\boldsymbol{\epsilon}_{\parallel}\neq\boldsymbol{\epsilon}_{s}$) the two sets of eigenvalues are independent.   Thus when either $\lambda_1$ or $\lambda_2$ are zero, the Helmholtz equation instead reduces to the \emph{two dimensional} Laplace equation.  At first sight this seems peculiar because even though one degree of freedom has disappeared from the system ($\boldsymbol{\epsilon}_{\parallel}$ in no longer a full rank matrix), it doesn't seem that this implies that the wave depends on only one coordinate.  As we shall show in this paper, this apparent two dimensional variation of the field is really one--dimensional propagation in disguise.  
\par
In addition, as summarized in figure~\ref{fig:gyrotropy}, when one of the eigenvalues of $\boldsymbol{\epsilon}_{\parallel}$ changes sign, a gyrotropic medium can change from transparent (propagating solutions) to opaque (without propagating solutions) or vice versa, whereas an `ordinary' material always goes from elliptic to hyperbolic dispersion.  It has recently been shown that such opaque gyrotropic media support unidirectional edge--states when bounded by a perfect conductor~\cite{davoyan2013}.  In the first part of this paper we explain the emergence of unidirectional propagation as one of the eigenvalues of $\boldsymbol{\epsilon}_{\parallel}$ reduces to zero, in terms a reduction of the dependence of the wave to a single complex spatial coordinate.
%
%
\noindent
\section{Complex principal axes in Maxwell's equations\label{sec:complex_coordinates}}
\par
Anisotropy can have a very interesting effect on wave propagation (see e.g.~\cite{figotin2005}), because the refractive index depends on the direction of propagation.  We now show that when one of the eigenvalues of the Hermitian tensor $\boldsymbol{\epsilon}_{\parallel}$ (\ref{eq:gyrotropic-medium}) equals zero (i.e. the refractive index is zero in a complex direction), the electromagnetic field becomes an analytic function of a single complex variable, with the one--dimensional Helmholtz equation determining its two dimensional variation in space.  To do this we return to Maxwell's equations to illustrate the behavior of all the field components in a gyrotropic medium.  We work in terms of the complex eigenvectors of $\boldsymbol{\epsilon}_{\parallel}$
\begin{equation}			
\boldsymbol{\epsilon}_{\parallel}\cdot\boldsymbol{\nabla}z^{\star}_{i}=\lambda_{i}\boldsymbol{\nabla}z^{\star}_{i}\label{eq:eigs}
\end{equation}
where these eigenvectors have been written as gradients of a pair of complex coordinates $z_{i}^{\star}$, obeying $\boldsymbol{\nabla}z_{i}\cdot\boldsymbol{\nabla}z_{j}^{\star}=\delta_{ij}$.  The physical meaning of a complex versus real set of basis vectors (or coordinates) can be understood in the same way as the relation between linear and circular polarization; a complex unit vector rotates in time, and thus includes part of the time dependence of the field.  Expanding the electric field in this basis
\begin{equation}
	\boldsymbol{E}=(\boldsymbol{\nabla}z_{1}^{\star}) E_{1}+(\boldsymbol{\nabla}z_{2}^{\star}) E_{2}\label{eq:electric_field_expansion}
\end{equation}
and the gradient operator in the conjugate basis
\begin{equation}
	\boldsymbol{\nabla}=(\boldsymbol{\nabla}z_{1})\frac{\partial}{\partial z_{1}}+(\boldsymbol{\nabla}z_{2})\frac{\partial}{\partial z_{2}}\label{eq:gradient_expansion}
\end{equation}
we find the first Maxwell equation $\boldsymbol{\nabla}\cdot\boldsymbol{\epsilon}_{\parallel}\cdot\boldsymbol{E}=0$ takes a rather simple form
\begin{equation}
	\lambda_{1}\frac{\partial E_{1}}{\partial z_{1}}+\lambda_{2}\frac{\partial E_{2}}{\partial z_{2}}=0.\label{eq:maxwell-1}
\end{equation}
which already shows that when one of the eigenvalues $\lambda_{i}$ of $\boldsymbol{\epsilon}_{\parallel}$ equals zero, one of the electric field components becomes a function of a single complex variable, $z_{i}$.  To see how the remaining field components behave in this limit we express the other Maxwell equations in the same basis, finding
\begin{align}
	\frac{\partial \eta_{0} H}{\partial z_{1}}&={\rm i}k_{0}\lambda_{2}E_{2}\nonumber\\
    \frac{\partial \eta_{0} H}{\partial z_{2}}&=-{\rm i}k_{0}\lambda_{1}E_{1}\nonumber\\
  \frac{\partial E_{2}}{\partial z_{1}^{\star}}-\frac{\partial E_{1}}{\partial z_{2}^{\star}}&={\rm i}k_{0}\eta_{0}H\label{eq:maxwell-2}
\end{align}
where $\eta_{0}=\sqrt{\mu_{0}/\epsilon_{0}}$, and the phase of the two complex coordinates has been chosen such that $\hat{\boldsymbol{z}}\times(\boldsymbol{\nabla}z_{1})=(\boldsymbol{\nabla}z_{2}^{\star})$.  Because the two complex coordinates $z_{1}$ and $z_{2}$ can both be written as linear combinations of $x$ and $y$, we can write e.g. $z_{2}$ as a linear combination of $z_{1}$ and $z_{1}^{\star}$.  A function that is independent of $z_{2}$ is thus equivalent to one that is independent of $z_{1}^{\star}$.  Therefore when one of the eigenvalues (we take this to be $\lambda_{1}$) equals zero, two of Maxwell's equations (\ref{eq:maxwell-1}--\ref{eq:maxwell-2}) reduce to the Cauchy--Riemann conditions~\cite{priestly2003}
\begin{align}
	\frac{\partial E_{2}}{\partial z_{1}^{\star}}&=0\qquad(\lambda_{1}=0)\nonumber\\
    \frac{\partial H}{\partial z_{1}^{\star}}&=0\label{eq:maxwell-3}
\end{align}
implying that for this range of material parameters, $E_2$ and $H$ are analytic functions of the single complex variable $z_1$.  Of the two remaining Maxwell equations, one states that $E_2$ is proportional to the derivative of $H$
\begin{equation}
	E_{2}(z_{1})=-\frac{{\rm i}\eta_0}{k_{0}\lambda_{2}}\frac{d H(z_{1})}{d z_{1}}\qquad(\lambda_1=0)\label{eq:maxwell-4}
\end{equation}
and the form of the field is entirely determined by the final Maxwell equation (\ref{eq:maxwell-2}), which reduces to
\begin{equation}
	\left(\frac{\partial z_{1}}{\partial z_{1}^{\star}}\right)_{z_{2}^{\star}}\frac{d E_{2}(z_1)}{d z_{1}}-\left(\frac{\partial E_{1}}{\partial z_{2}^{\star}}\right)_{z_{1}^{\star}}={\rm i}k_{0}\eta_{0}H(z_1).\label{eq:maxwell-5}
\end{equation}
(the subscripts have been introduced to avoid ambiguity about what is being held constant during differentiation).
\par
We cannot make use of equation (\ref{eq:maxwell-5}) until we know something about the field component $E_{1}$, which the remaining Maxwell equations (\ref{eq:maxwell-3}--\ref{eq:maxwell-4}) do not tell us.  Due to the dependence of $E_2$ and $H$ on $z_1$ it must be the case that $(\partial E_{1}/\partial z_{2}^{\star})_{z_{1}^{\star}}$ is also a function of only $z_1$.  To satisfy this requirement we assume that $E_{1}$ itself is a function of $z_1$, an assumption that---for the cases investigated here---only holds when the region within the cavity is simply connected (see following section).  To fix the functional form of $E_{1}$ we define a meromorphic function $F$ as the ratio between the two in--plane electric field components
\begin{equation}
	F(z_1)=\frac{E_{1}(z_1)}{E_{2}(z_1)}\label{eq:Fdef}
\end{equation}
This function $F$ can be found from the boundary conditions on the electric field.  For instance, if the gyrotropic medium is enclosed by a perfect conductor then the in plane electric field must vanish at its surface.  If the perfect conductor follows the curve $z_1(\sigma)$ then this is equivalent to
\begin{equation}
	\hat{\boldsymbol{n}}(\sigma)\cdot\left(\boldsymbol{\nabla}z_{1}^{\star}\right)E_{1}(z_1(\sigma))=-\hat{\boldsymbol{n}}(\sigma)\cdot\left(\boldsymbol{\nabla}z_{2}^{\star}\right)E_{2}(z_1(\sigma))\label{eq:perfect-conductor}
\end{equation}
where $\hat{\boldsymbol{n}}(\sigma)$ is the tangent to the curve.  Rearranging (\ref{eq:perfect-conductor}) to give the value of the function $F$ (\ref{eq:Fdef}) on the boundary, we find that $F$ is the meromorphic function that reduces to
\begin{equation}
F(z_1(\sigma))=\frac{\hat{\boldsymbol{n}}\cdot\boldsymbol{\nabla}z_{2}^{\star}}{\hat{\boldsymbol{n}}\cdot\boldsymbol{\nabla}z_{1}^{\star}}\label{eq:Fdef}
\end{equation}
on the closed curve $z_1(\sigma)$.
\par
Assuming the cavity is simply connected, and we can find this function $F$, our final Maxwell equation (\ref{eq:maxwell-5}) reduces to a \emph{one dimensional} Helmholtz equation
\begin{equation}
	\frac{d}{d z_{1}}\left[\left(\frac{\partial z_{1}}{\partial z_{1}^{\star}}\right)_{z_{2}^{\star}}-\left(\frac{\partial z_{1}}{\partial z_{2}^{\star}}\right)_{z_{1}^{\star}}F(z_{1})\right]\frac{d H(z_{1})}{d z_{1}}+\lambda_{2}k_0^{2} H(z_1)=0\label{eq:1dhelm}
\end{equation}
where equation (\ref{eq:maxwell-4}) was used to eliminate $E_2$.  Rather surprisingly, equation (\ref{eq:1dhelm})---which determines the \emph{two}--dimensional variation of the magnetic field within a closed cavity---is of exactly the same form as the one dimensional Helmholtz equation for an electromagnetic wave propagating through a planar medium.  This similarity becomes exact when the one--dimensional Helmholtz equation is analytically continued to complex values of the spatial coordinates (as in~\cite{horsley2016}).  For instance, in a planar medium the magnetic field of an $H$--polarized wave propagates according to $(d/dx)\epsilon^{-1}(x)d H/dx + k_0^2\mu(x)H(x)=0$~\cite{born2003}, so that the spatial variation of $F(z_1)$ (here determined by the gyrotropic medium and the cavity shape) is analogous to the spatial variation of an effective permittivity in 1D.
\par
As mentioned in the introduction, provided the cavity is simply connected, the reduction of the field to an analytic function of a single complex variable automatically implies unidirectional propagation, simply because the field component $H$ is a continuous function without singularities.  The field therefore has a Taylor expansion in positive powers of $z_1$; e.g. $H=\sum_{n\geq0}H_n (x+{\rm i}y)^n=\sum_{n\geq0}H_n |r|^n\exp({\rm i}n\theta)$, and thus angular momenta of only one sign.  This is evident in figure~\ref{fig:gyrotropy}b, where as the determinant of $\boldsymbol{\epsilon}_{\parallel}$ is reduced to zero the lowest cavity mode acquires a phase variation that winds only in a clockwise sense.
%
%
\noindent
\section{Unidirectional states in a cylindrical cavity\label{sec:example}}
\begin{figure}[ht!]
	\includegraphics[width=16cm]{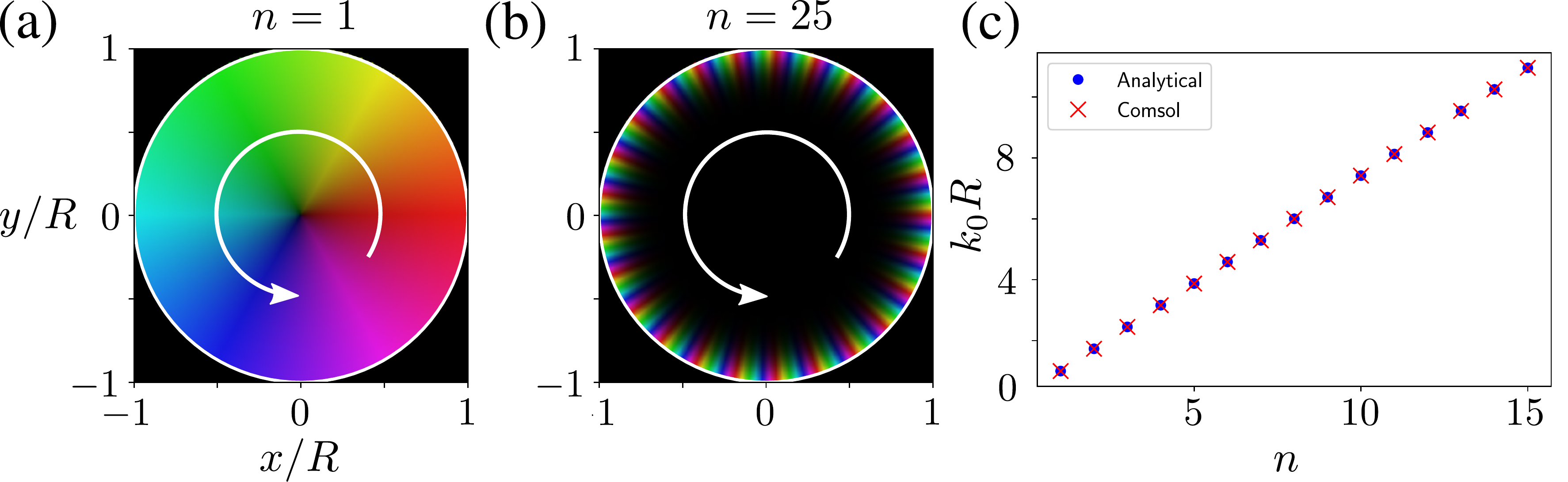}
	\caption{In a medium with a permittivity tensor given by (\ref{eq:permittivity_example}) (in this example the parameter $\lambda$ has the value $\lambda=4$) the spatial dependence of the out of plane magnetic field is simply a power of the complex coordinate $z=x+{\rm i}y$.  The first two panels show the magnetic field for the $n^{\rm th}$ cavity mode, with (a) $n=1$ and (b) $n=25$, illustrating the appearance of edge states in the large $n$ limit.  Panel (c) shows a comparison of the cavity eigenfrequencies computed numerically, and from the analytical expression (\ref{eq:z_freq}).\label{fig:cavity}}
\end{figure}
\par
Despite the reduction to a one--dimensional dependence, it is difficult to find exact solutions to the Helmholtz equation (\ref{eq:1dhelm}) for arbitrary gyrotropic media filling arbitrarily shaped cavities.  Here we explore a simple example which illustrates the essential features of the results given above: in a simply connected cavity, when one of the eigenvalues of $\boldsymbol{\epsilon}_{\parallel}$ is zero, then the allowed modes depend on only one complex coordinate, and can therefore propagate in only one direction.  
\par
First we use our previous results to design a permittivity where the electromagnetic field is an analytic function of $z=x+{\rm i}y$ and thus propagates only counter--clockwise.  The two complex coordinates are determined by the conditions $\boldsymbol{\nabla}z_{i}\cdot\boldsymbol{\nabla}z_{j}^{\star}=\delta_{ij}$  and $\hat{\boldsymbol{z}}\times\boldsymbol{\nabla}z_{1}=\boldsymbol{\nabla}z_{2}^{\star}$, and we choose
\begin{align}
	z_{1}&=\frac{1}{\sqrt{2}}(x+{\rm i}y)\nonumber\\
    z_{2}&=\frac{{\rm i}}{\sqrt{2}}(x-{\rm i}y).\label{eq:z1z2}
\end{align}
Given the desired dependence of the field on $z_1$ we fix the eigenvalues as $\lambda_1=0$ and $\lambda_{2}=\lambda$.  Using our definition of the permittivity tensor (\ref{eq:eigs}), we find that the requisite $\boldsymbol{\epsilon}_{\parallel}$ is given by
\begin{equation}	\boldsymbol{\epsilon}_{\parallel}=\lambda(\boldsymbol{\nabla}z_{2}^{\star})\otimes(\boldsymbol{\nabla}z_{2})=\frac{\lambda}{2}\left(\begin{matrix}1&-{\rm i}\\{\rm i}&1\end{matrix}\right)\label{eq:permittivity_example}.
\end{equation}
For a perfectly conducting circular boundary of radius $R$, the coordinates of the boundary are $x(\sigma)=R\cos(\sigma)$ and $y(\sigma)=R\sin(\sigma)$ ($\sigma\in[-\pi,\pi]$), with the tangent to the boundary given by $\hat{\boldsymbol{n}}(\sigma)=\cos(\sigma)\hat{\boldsymbol{y}}-\sin(\sigma)\hat{\boldsymbol{x}}$.  Thus the function $F$ appearing in (\ref{eq:1dhelm}) must satisfy the boundary condition (\ref{eq:Fdef}) at $|z_1|=R/\sqrt{2}$
\begin{equation}
	F(R {\rm e}^{{\rm i}\sigma}/\sqrt{2})=-\frac{\hat{\boldsymbol{n}}\cdot(\boldsymbol{\nabla}z_{2}^{\star})}{\hat{\boldsymbol{n}}\cdot(\boldsymbol{\nabla}z_{1}^{\star})}=-{\rm i}{\rm e}^{2{\rm i}\sigma}\label{eq:bc_cylinder}
\end{equation}
which uniquely fixes the functional form of $F(z_1)$ within the cavity to be $F(z_1)=-{\rm i}(\sqrt{2}z_1/R)^{2}$.  Having found the form of $F$, the Helmholtz equation for the out of plane magnetic field (\ref{eq:1dhelm}) takes the form
\begin{equation}
	\frac{d}{d z_{1}}\left(\frac{z_{1}}{R}\right)^{2}\frac{d H(z_{1})}{d z_{1}}-\frac{\lambda k_0^{2}}{2} H(z_1)=0
\end{equation}
the solution to which is a power of $z_1$,
\begin{equation}
	H_n=H_0\left(\frac{\sqrt{2}z_{1}}{R}\right)^{n}=H_0\left(\frac{r}{R}\right)^{n}{\rm e}^{{\rm i}n\theta}\label{eq:Hn}
\end{equation}
($H_0$ is a constant) with the eigenfrequencies given by
\begin{equation}
	k_{0,n}=\frac{\sqrt{2n(n+1)}}{\sqrt{\lambda} R}\label{eq:z_freq}
\end{equation}
For real $\lambda$ the eigenfrequencies $k_0$ are real and the spectrum is the same for positive and negative $n$ (with a degeneracy at zero frequency of the $n=0$ and $n=-1$ modes).  However the fact that the field amplitudes should remain finite throughout the cavity restricts $n$ to positive values, and thus fixes the propagation to be in the anti--clockwise sense for all cavity modes.  Figure~\ref{fig:cavity}c shows that the frequencies (\ref{eq:z_freq}) can be reproduced numerically from a finite--element simulation.  In the limit of large $n$ the wave amplitude (\ref{eq:Hn}) is close to zero except at the boundary of the cavity and thus appears as a unidirectional edge state (see figure~\ref{fig:cavity}b).  Note that this is true even in the non--Hermitian case where $\lambda$ is complex (the system has loss or gain), it is just that each cavity mode is amplified or diminished over time.
\par
From this example it is thus evident that one can design materials supporting unidirectional edge--states through specifying a complex valued principal axis where the refractive index is zero.  We shall return to this point in the final section, showing that transformation optics can be generalized to perform complex valued rotations, transforming `ordinary' anisotropic media into gyrotropic media supporting one--way propagation.
%
%
\section{Sensitivity to the cavity topology}
\par
Because the wave reduces to a function of a single complex variable in the limit ${\rm det}[\boldsymbol{\epsilon}_{\parallel}]\to0$, the field inherits the same sensitivity to the topology of the cavity as an analytic function within a portion of the complex plane, and this can lead to some quite unusual physical effects.  For example, the uniqueness theorem~\cite{priestly2003} guarantees that if an analytic function vanishes in a region of the complex plane that is bigger than a point then it vanishes everywhere.  Thus if the cavity contains any object which forces the out of plane magnetic field to zero, then the field is zero throughout the cavity, and no modes can be supported.  Similarly, if a function $f(z)$ is analytic in a region of the complex plane excluding a hole centred on $z_0$ then its expansion $f(z)=\sum_{n}f_n(z-z_0)^n$ can include both positive and negative $n$.  Thus when our cavity contains a hole, in general the propagation will cease to be unidirectional (which is because waves can be bound to the outer boundaries of the holes, and run in the opposite sense compared to those bound to the inner boundary of the outermost conductor).  In this section we use the example of a cylindrical cavity to illustrate both of these effects. 
%
%
\begin{figure}[h!]
	\includegraphics[width=16cm]{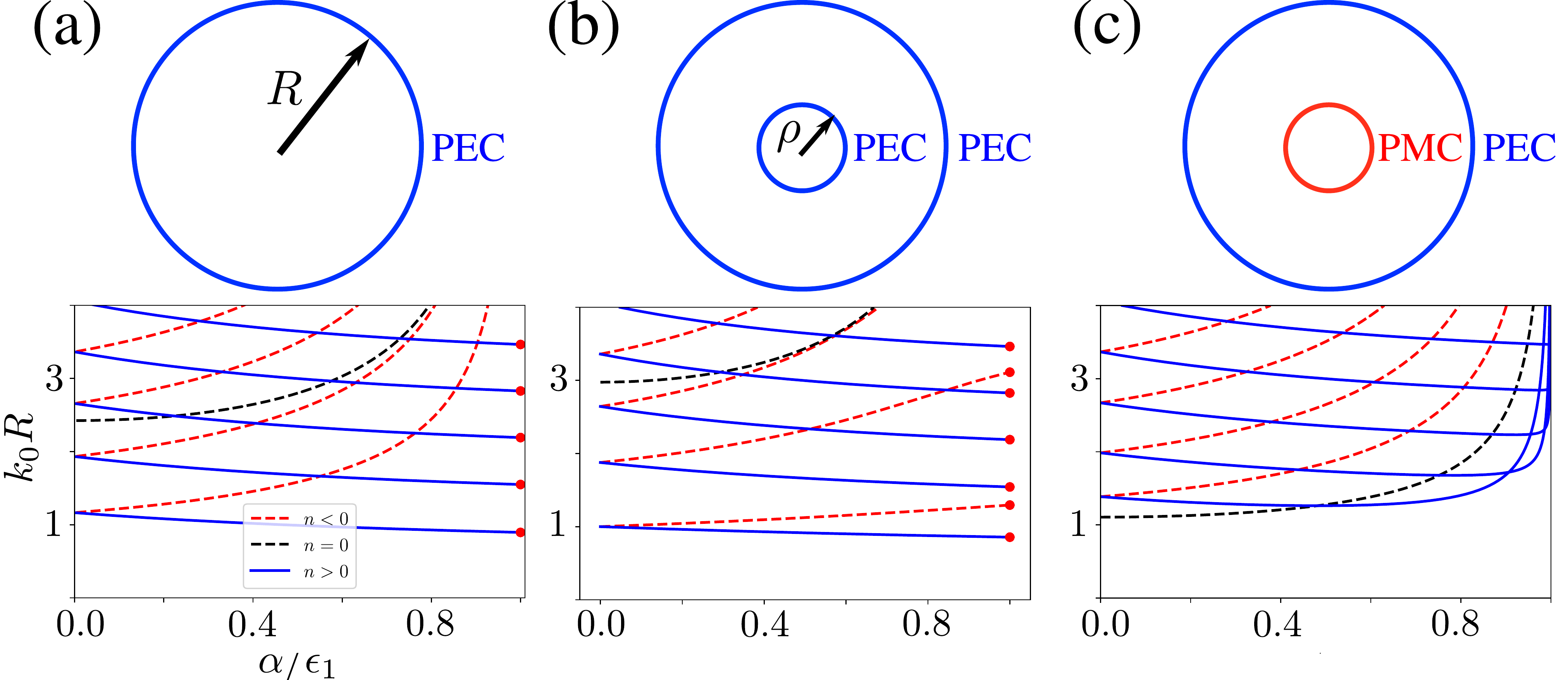}
	\caption{The behaviour of the cavity modes in the limit $\lambda_1=\epsilon_1-\alpha\to0$ depends strongly on the connectedness of the gyrotropic medium filling the cavity. (a) A simply connected cavity. (b) A cavity containing a perfectly conducting cylinder of radius  $\rho=0.3R$. (c) A cavity containing a perfect magnetic conductor of the same radius.\label{fig:topology_effect}}
\end{figure}
\par
We consider the three cases shown in the upper half of figure~\ref{fig:topology_effect}, where the cylindrical cavity is either (a) simply connected; (b) contains a perfectly electrically conducting cylinder of radius $\rho$; or (c) contains a perfectly magnetically conducting cylinder of the same radius.  We consider a cylindrical cavity containing a medium with the following in--plane permittivity tensor (c.f.~\cite{davoyan2013})
\begin{align}
\boldsymbol{\epsilon}_{\parallel}&=\lambda_1(\boldsymbol{\nabla}z_{1}^{\star})\otimes(\boldsymbol{\nabla}z_{1})+\lambda_2(\boldsymbol{\nabla}z_{2}^{\star})\otimes(\boldsymbol{\nabla}z_{2})\nonumber\\[10pt]
    &=\frac{1}{2}\left[(\epsilon_{1}-\alpha)(\boldsymbol{\nabla}z^{\star})\otimes(\boldsymbol{\nabla}z)+(\epsilon_{1}+\alpha)(\boldsymbol{\nabla}z)\otimes(\boldsymbol{\nabla}z^{\star})\right]\label{eq:perm_lambda}
\end{align}
where $z_{1}$ and $z_{2}$ are defined as in (\ref{eq:z1z2}), and $z=x+{\rm i}y$.  The frequencies of the cavity modes as a function of $\alpha/\epsilon_{1}$ ($\lambda_{1}=0$ when $\alpha/\epsilon_{1}=1$) are shown in the lower three panels of figure~\ref{fig:topology_effect}, with blue solid lines for the positive angular momentum ($n>0$) modes, and dashed lines for the $n\leq0$ modes.  The frequencies were calculated semi--analytically (see appendix), with the red dots showing the analytical predictions for the limit $\lambda_1\to0$.
\par
In the case of a simply connected cavity (figure~\ref{fig:topology_effect}a)---already treated in the previous section---we expect that as $\lambda_{1}\to0$, only modes of positive angular momenta should remain.  We see that this is the case, with the solid blue lines approaching the frequencies (\ref{eq:z_freq}) shown as red dots, and the dashed red lines diverging to infinite frequency.  By contrast, figure~\ref{fig:topology_effect}c shows that when a cylinder of perfect magnetic conductor is placed within the cavity, the frequencies of all the modes diverge as $\lambda_{1}\to 0$ (although the sensitivity to the material parameters is more for the higher order modes).  This agrees with our prediction based on the uniqueness theorem of complex analysis.
\par
Figure~\ref{fig:topology_effect}b shows that when a cylinder of perfect electrical conductor is placed within the cavity, then cavity modes of both positive and negative angular momenta are allowed in the limit $\lambda_1\to0$.  This agrees with our prediction based on the behaviour of an analytic function, defined in a region of the complex plane excluding a hole.  We now briefly show how this behaviour emerges, using a similar argument to the previous section.
\par
In the limit $\epsilon_{1}\to\alpha$, the permittivity  (\ref{eq:perm_lambda}) reduces to that of section~\ref{sec:example}, and we can use the same system of coordinates $z_{1},z_{2}$ we did there, applying a slight generalization of the argument.  Our field components $H$ and $E_2$ must still reduce to analytic functions of $z_1$, even in the presence of the conducting cylinder.  For a fixed angular momentum they must therefore be given by powers of $z_1$ as in equation (\ref{eq:Hn})
\begin{align*}
	H(z_1)&=H_0 \left(\frac{\sqrt{2}z_1}{R}\right)^{n}\\
    E_{2}(z_1)&=\frac{\eta_0\sqrt{2} H_0}{{\rm i}k_0 R\lambda_2}n \left(\frac{\sqrt{2}z_1}{R}\right)^{n-1}
\end{align*}
where the electric and magnetic field are related by (\ref{eq:maxwell-4}).  The boundary conditions on the inner and outer cylinder are identical to one another, and are given by (\ref{eq:bc_cylinder}) 
\begin{equation}
	F(R{\rm e}^{{\rm i}\sigma}/\sqrt{2})=F(\rho{\rm e}^{{\rm i}\sigma}/\sqrt{2})=-{\rm i}{\rm e}^{2{\rm i}\sigma}\label{eq:two_bcs}
\end{equation}
It is impossible to satisfy these two boundary conditions with $F$ depending on only $z_1$.  Instead it must be a function of both $z_1$ and $z_1^{\star}$.  Given the form of the Helmholtz equation (\ref{eq:maxwell-5}), the only possibility is that electric field component $E_1$ is a sum of a power of $z_1$ and a power of $z_1^{\star}$.  After a little consideration we find that $E_{1}$ must have the following form
\begin{equation}
	E_1(z_1,z^{\star}_1)=-\frac{\sqrt{2}\eta_0 H_0}{k_0 R\lambda_2}n \left[\kappa_{n} \left(\frac{\sqrt{2}z_1}{R}\right)^{n+1}+(1-\kappa_{n})\left(\frac{\sqrt{2}z_1^{\star}}{R}\right)^{-n-1}\right]\label{eq:E1}
\end{equation}
with $\kappa_{n}$ a constant.  Imposing the condition (\ref{eq:two_bcs}) and the definition of $F$ as a ratio of field components, this fixes the value of the unknown $\kappa_{n}$
\[
	\kappa_n=\frac{1-\left(\frac{\rho}{R}\right)^{-2n}}{\left(\frac{\rho}{R}\right)^{2}-\left(\frac{\rho}{R}\right)^{-2n}}
\]
Substituting (\ref{eq:E1}) into (\ref{eq:maxwell-5}) then gives us the frequencies of the cavity eigenmodes
\begin{equation}
	k_0=\frac{\sqrt{2\kappa_n n(n+1)}}{\sqrt{\lambda_2} R}\label{eq:cavity_hole}
\end{equation}
The frequencies given by (\ref{eq:cavity_hole}) are shown as the red dots in figure~\ref{fig:topology_effect}b, demonstrating that the numerical results do indeed approach these values as $\lambda_1\to0$~\footnote{Note that when $n=-1$ the numerator of (\ref{eq:cavity_hole}) becomes indeterminate and must be evaluated as a limit, giving $\lim_{n\to-1}(n+1)\kappa_n=[1-(\rho/R)^2]/[2(\rho/R)^2\ln(\rho/R)]$}.  Because of the excluded region in the centre of the cavity there is no argument to exclude the negative $n$ modes, and as shown in the figure these remain in the limit.  Changing the topology of the cavity thus changes whether the supported modes are uni--directional or not, and this can be understood entirely on the basis of the theory of complex functions.  
\par
We note also that although it is not possible to manufacture materials with sufficient precision to say that the determinant of $\boldsymbol{\epsilon}_{\parallel}$ is exactly zero, the numerical results given in this section and summarized in figure~\ref{fig:topology_effect} show that the predicted behaviour remains for at least a band of frequencies, for values of the permittivity where the determinant is only close to zero.  

%
%
\section{Transformation optics and unidirectional propagation}
\par
To conclude, we shall develop the intuition we have been using so far: that of uni--directional wave propagation in terms of materials with principal axes pointing in complex directions.  We show how a complex valued rotation of the coordinate system can convert a zero index (and in general hyperbolic) material into one supporting one--way edge states.  This is akin to transformation optics, which establishes the equivalence between a change in material parameters and a change in coordinate system~\cite{pendry1996,pendry2006,leonhardt2010}.
\par
Take the simplest case of transformation optics where the medium is planar and homogeneous.  Assuming the same polarization as in the previous sections, we expand the fields and the derivatives in terms of an arbitrary complex orthonormal basis $\boldsymbol{e}_{i}$ (rather than the basis determined by the in--plane permittivity (\ref{eq:eigs})) where
\begin{align}
	\boldsymbol{e}_{i}\cdot\boldsymbol{e}_{j}^{\star}&=\delta_{ij}\nonumber\\
    \boldsymbol{e}_{1}\times\boldsymbol{e}_{2}&=\hat{\boldsymbol{z}}
\end{align}
We do not here consider the complication of letting this basis depend on position.  This complex basis also defines a set of complex coordinates that we can infer from the gradient operator
\[
	\frac{\partial}{\partial z_{i}}=\boldsymbol{e}_{i}^{\star}\cdot\boldsymbol{\nabla}=\frac{\partial x}{\partial z_{i}}\frac{\partial}{\partial x}+\frac{\partial y}{\partial z_{i}}\frac{\partial}{\partial y}
\]
and similarly a conjugate set of coordinates
\[
	\frac{\partial}{\partial z_{i}^{\star}}=\boldsymbol{e}_{i}\cdot\boldsymbol{\nabla}=\frac{\partial x}{\partial z_{i}^{\star}}\frac{\partial}{\partial x}+\frac{\partial y}{\partial z_{i}^{\star}}\frac{\partial}{\partial y}.
\]
Expressing Maxwell's equations in terms of these basis vectors we find that they take a similar form to (\ref{eq:maxwell-1}--\ref{eq:maxwell-3})
\begin{align}
	e_{ij}\frac{\partial E_{j}}{\partial z_{i}}&={\rm i}\omega\mu_{0}H\nonumber\\
    e_{i j}\frac{\partial H}{\partial z_{j}^{\star}}&=-{\rm i}\omega\epsilon_{0}\epsilon_{i j}E_{j}\label{eq:maxwell_complex}
\end{align}
where the components of the in--plane permittivity tensor are $\epsilon_{ij}=\boldsymbol{e}_{i}^{\star}\cdot\boldsymbol{\epsilon}\cdot\boldsymbol{e}_{j}$ and $e_{i j}$ is the antisymmetric unit tensor with components $e_{11}=e_{22}=0$ and $e_{12}=-e_{21}=1$.  The two equations given in (\ref{eq:maxwell_complex}) are the general expression for the Maxwell equations in this class of two dimensional complex coordinate systems.  We note that this form of the equations is a kind of generalization of the coordinates used in optical conformal mapping~\cite{leonhardt2006}, although we do not consider conformal transformations here.
\begin{figure}
	\includegraphics[width=16cm]{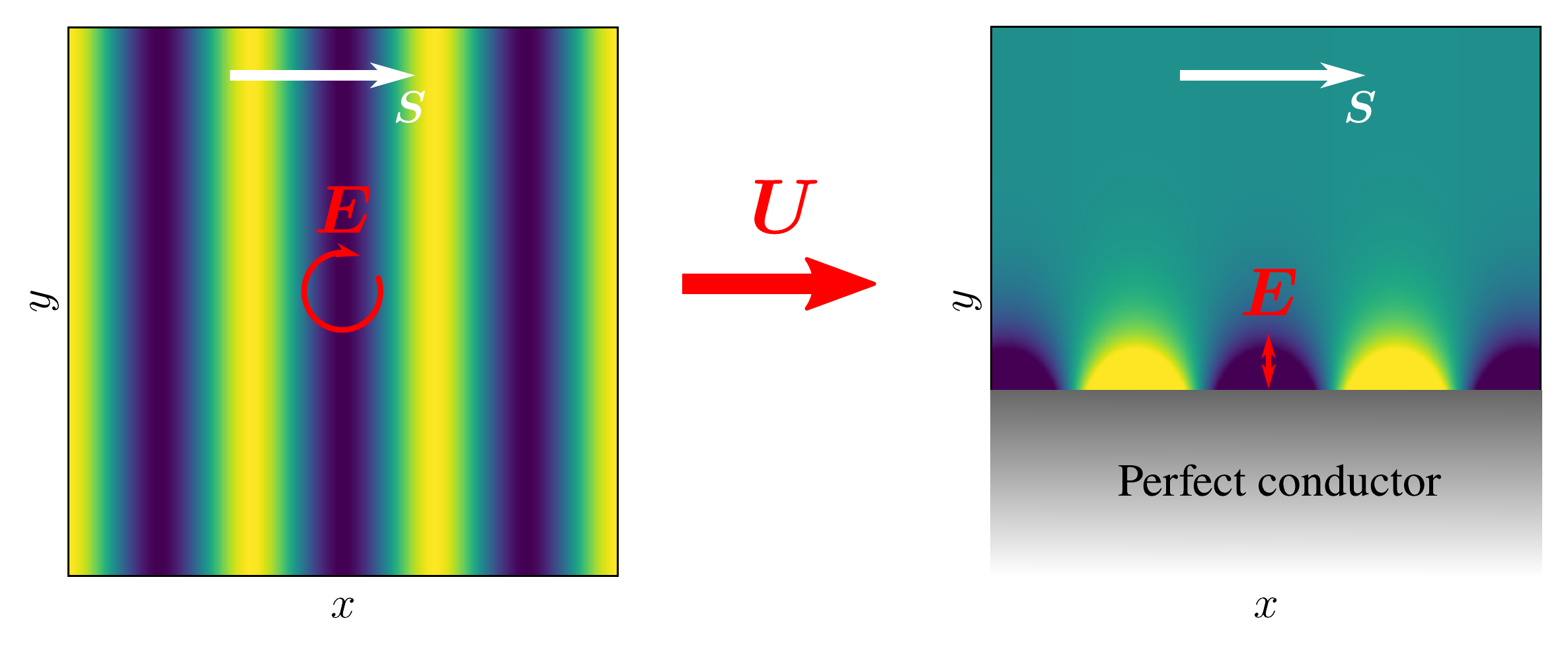}
    \caption{Maxwell's equations take the same form in a class of complex coordinate bases.  A change of basis is equivalent to a change in material properties.  The above schematic example shows the effect of the unitary transformation $\boldsymbol{U}$ (\ref{eq:coord_trans}), where a plane wave in a hyperbolic anisotropic medium is transformed to a one--way edge state via medium (\ref{eq:trans_perm}).  The vector $\boldsymbol{S}$ indicates the direction of power flow in these two cases.\label{fig:complex_rotation}}
\end{figure}
\par
As the Maxwell equations (\ref{eq:maxwell_complex}) take the same form in all such coordinate systems, we can take the point of view of transformation optics, considering a unitary transformation of the basis $\boldsymbol{e}_{i}=U_{ij}^{\star}\bar{\boldsymbol{e}}_{j}$, and consequent change in coordinates.  This transformation can either be viewed as a change in perspective (viewing the same fixed field from a different set of coordinates), or as fixed perspective where there has been a change in the field components, brought about by the following change in the in--plane permittivity tensor
\begin{equation}
	\bar{\epsilon}_{ij}=U^{\dagger}_{i k} \epsilon_{k m} U_{m j}.\label{eq:complex_transformation}
\end{equation}
\par
As an example consider the case described to in the previous sections; we start with an anisotropic medium with the refractive index zero along a fixed direction in space, rotating the coordinate system so that the directions becomes complex.  We consider a diagonal permittivity tensor
\begin{equation}
	\boldsymbol{\epsilon}_{\parallel}=\lambda\boldsymbol{e}_{2}\otimes\boldsymbol{e}_{2}^{\star}=\left(\begin{matrix}0&0\\0&\lambda\end{matrix}\right)\label{eq:example_medium}
\end{equation}
where
\begin{align}
	\boldsymbol{e}_{1}&=\boldsymbol{\nabla}z_{1}=\frac{\hat{\boldsymbol{x}}+{\rm i}\eta\hat{\boldsymbol{y}}}{\sqrt{1+\eta^{2}}}\nonumber\\
    \boldsymbol{e}_{2}&=\boldsymbol{\nabla}z_{2}=\frac{\hat{\boldsymbol{y}}+{\rm i}\eta\hat{\boldsymbol{x}}}{\sqrt{1+\eta^{2}}}\label{eq:basis_example}
\end{align}
where we have introduced the quantity $\eta$ to keep track of the distinction between the coordinates and their conjugates in (\ref{eq:maxwell_complex}).  In the limit $\eta\to0$ the medium (\ref{eq:example_medium}) reduces to an anisotropic medium with zero index for propagation along the $y$--axis.  One solution for the field in such a medium is given by
\begin{align}
	H&=H_0{\rm e}^{{\rm i}k_1 z_{1}^{\star}}\nonumber\\
    \boldsymbol{E}&=\frac{k_{1}\eta_0 H_0}{k_0\lambda}{\rm e}^{{\rm i}k_1 z_{1}^{\star}}(\boldsymbol{e}_{2}+{\rm i}\boldsymbol{e}_{1})\label{eq:pw_anisotropic}
\end{align}
As we saw in section~\ref{sec:complex_coordinates}, zero index in a particular direction means that one of the field components (in this case $E_1$) is undetermined, and must be fixed by the boundary conditions.  Here we choose $E_{1}={\rm i}E_{2}$, so that the electric field of the plane wave (\ref{eq:pw_anisotropic}) rotates in the $x$-$y$ plane (see figure~\ref{fig:complex_rotation}).  Substituting (\ref{eq:pw_anisotropic}) into (\ref{eq:maxwell_complex}), we find the dispersion relation connecting $k_1$ to $k_0$ is
\begin{equation}
	k_1^{2}=\frac{1+\eta^{2}}{1-2\eta-\eta^{2}}\lambda k_{0}^{2}\label{eq:complex_dispersion}
\end{equation}
which reduces to the usual dispersion relation $k_1=\pm\sqrt{\lambda}k_{0}$ for an anisotropic medium when $\eta\to0$, propagation being allowed in both directions.  Now suppose we describe this field in a different coordinate system
\begin{equation}
	z_{i}=U^{\star}_{ij}\bar{z}_{j}\to\left(\begin{matrix}z_{1}\\z_{2}\end{matrix}\right)=\frac{1}{\sqrt{2}}\left(\begin{matrix}1&-{\rm i}\\-{\rm i}&1\end{matrix}\right)\left(\begin{matrix}\bar{z}_{1}\\\bar{z}_{2}\end{matrix}\right)\label{eq:coord_trans}
\end{equation}
(the basis vectors (\ref{eq:basis_example}) transform in the same way).  In this basis the electric and magnetic fields are given by
\begin{align}
	H&=H_0{\rm e}^{{\rm i}\frac{k_1}{\sqrt{2}}(\bar{z}_{1}^{\star}+{\rm i}\bar{z}_{2}^{\star})}\nonumber\\
    \boldsymbol{E}&=\frac{\sqrt{2}k_{1}\eta_0 H_0}{k_0\lambda}{\rm e}^{{\rm i}\frac{k_1}{\sqrt{2}}(\bar{z}_{1}^{\star}+{\rm i}\bar{z}_{2}^{\star})}\bar{\boldsymbol{e}}_{2}\label{eq:transformed_fields}
\end{align}
which is also a solution to Maxwell's equations, with the permittivity in the new coordinate system $\bar{z}_{1}$, $\bar{z}_{2}$ given by
\begin{equation}
	\boldsymbol{\epsilon}_{\parallel}=\frac{\lambda}{2}\left(\begin{matrix}1&-{\rm i}\\{\rm i}&1\end{matrix}\right)\label{eq:trans_perm}
\end{equation}
which is the same form of the permittivity (\ref{eq:permittivity_example}) we constructed to support uni--directional propagation in section~\ref{sec:example} (and that considered in the recent work of Davoyan, Engheta and Silveirinha~\cite{davoyan2013,silveirinha2015}).  We now take the point of view of transformation optics and interpret the change in permittivity tensor from (\ref{eq:example_medium}) to (\ref{eq:trans_perm}) as that which induces the change in the field from (\ref{eq:pw_anisotropic}) to (\ref{eq:transformed_fields}) with the basis vectors and coordinates remaining fixed.  Taking the limit $\eta\to 0$, the basis vectors of our original system (\ref{eq:basis_example}) become as close to $\hat{\boldsymbol{x}}$ and $\hat{\boldsymbol{y}}$ as we like, and the coordinates become arbitrarily close to a Cartesian system.  In this system, the field exponentially decays along the $x$--axis, and the $\hat{\boldsymbol{x}}$ component of the electric field vanishes
\begin{align}
	H&\to H_0{\rm e}^{{\rm i}\frac{k_1}{\sqrt{2}}(x+{\rm i}y)}\nonumber\\
    \boldsymbol{E}&\to\frac{\sqrt{2}k_{1}\eta_0 H_0}{k_0\lambda}{\rm e}^{{\rm i}\frac{k_1}{\sqrt{2}}(x+{\rm i}y)}\hat{\boldsymbol{y}}\label{eq:transformed_fields2}
\end{align}
For $k_1>0$ this solution (\ref{eq:transformed_fields2}) describes a unidirectional state that can only propagate to the right, bound to a perfect conductor which exists in the region $y<0$. For $k_1<0$ it describes a unidirectional state propagating to the left, bound to a perfect conductor which exists in the region $y>0$.  For this complex rotation we are forced to have propagation in an anti--clockwise sense (just as we were in section~\ref{sec:example}), whatever the direction of propagation of the initial plane wave.  Therefore, waves trapped at the surface of a conductor and exhibiting uni--directional propagation can be understood as the complex rotation (\ref{eq:coord_trans}) of waves with a circulating electric field (\ref{eq:pw_anisotropic}) in a planar medium where the refractive index is zero along one of the axes.  This is shown schematically in figure~\ref{fig:complex_rotation}.  In the transformed system, the rotating field becomes a linear polarization, and the homogeneous plane wave becomes an inhomogeneous one, with the direction of decay vs. propagation being fixed by the parameters of the rotation.
\par
This completes the picture, where a gyrotropic medium exhibiting one--way edge states can be understood as inheriting its properties from a complex rotation of an equivalent medium with principal axes pointing along fixed directions in real space.  This transformation optics inspired approach may prove useful in more general situations, where---for instance---the initial medium is inhomogeneous in space.
%
%
\section{Summary and conclusions}
\par
We have investigated an alternative way to understand and design electromagnetic materials that exhibit uni--directional propagation, possessing edge--states that can propagate one way and are bound to the interface with a perfect conductor.  The central point to take away from the first three sections of this paper is that when the refractive index is zero in a fixed complex direction then the electromagnetic field in a planar medium becomes dependent on a single complex variable $z$.  If the medium does not contain any holes then the wave is an analytic function of position, obeying the Cauchy--Riemann conditions in real space.  This analyticity automatically guarantees that the wave can only propagate in one direction, because its Taylor expansion $H(z)=\sum_{n\geq 0}H_n(z)^{n}=\sum_{n\geq 0}H_n r^{n}\exp({\rm i}n\theta)$ contains only angular momentum of one sign.  As we have shown, one can thus design media supporting one--way edge states through simply ensuring that the field is an analytic function of a single complex variable.
\par
There are a number of immediate results that come from this insight.  Firstly, we know that although analytic functions of a complex variable $z=x+{\rm i}y$ exhibit a two dimensional variation in the complex plane, this is a kind of one--dimensional variation in disguise.  The same is true for the electromagnetic field, and we have shown that the field in these two dimensional materials obeys a one--dimensional Helmholtz equation (\ref{eq:1dhelm}) of the same form as governs an electromagnetic wave propagating in 1D.  Secondly, we know that the behaviour of an analytic function is very different depending on the connectedness of the region in which it is defined.  This is also true for the fields in a material with zero index in a complex direction.  Non--simply connected cavities no longer support uni--directional propagation, and in some cases support no propagation at all.
\par
Finally we showed how a complex rotation of the coordinates can be used to convert Maxwell's equations in ordinary anisotropic media into Maxwell's equations in gyrotropic media, where the consequent dependence on a pair of complex coordinates can be used to design materials where waves can be trapped at an interface with a perfect conductor.  This is a kind of extension of transformation optics that could prove useful in a wider context, to find new media supporting unidirectional propagation without having to compute Chern numbers.  The findings of this work may prove useful for understanding and designing new planar structures where the allowed modes have a pre--specified direction of propagation.  The well--known results of complex analysis can be a very powerful tool as part of such a design process.

\acknowledgements
SARH acknowledges financial support from a Royal Society TATA University Research Fellowship (RPG-2016-186).  He is very grateful for the suggestions of C. G. King and T. G. Philbin.
%
%
\appendix
\noindent
\section{Appendix: Calculation of cavity modes}
\par
To find the modes in a cylindrical cavity filled with a gyrotropic medium of the form (\ref{eq:perm_lambda}) we apply the choice of complex coordinates (\ref{eq:z1z2}) to Maxwell's equations (\ref{eq:maxwell-2}), finding the following equation governing the out of plane magnetic field
\begin{equation}
	4\frac{\partial^{2}H}{\partial z\partial z^{\star}}+k^{2}H=0\label{eq:H_gen_eq}
\end{equation}
where $k^2=(\epsilon_1^2-\alpha^2)k_0^2/\epsilon_1$, and $z=x+{\rm i}y$.  Note that this is the same as equation (\ref{eq:TMwveqn}), just written in terms of the complex variables $z$ and $z^{\star}$.  In general the electric field is given in terms of $H$ by
\[
	\boldsymbol{E}(z,z^{\star})=\frac{\eta_0}{k_0}\left[\frac{1}{\epsilon_{1}-\alpha}\frac{\partial H}{\partial z^{\star}}(\boldsymbol{\nabla}z^{\star})-\frac{1}{\epsilon_1+\alpha}\frac{\partial H}{\partial z}(\boldsymbol{\nabla}z)\right].
\]
Given the cylindrical symmetry of the system, the solution to (\ref{eq:H_gen_eq}) is given by a sum of Bessel functions
\begin{equation}
	H(z,z^{\star})=\left[J_n(k \sqrt{zz^{\star}})+a Y_n(k \sqrt{zz^{\star}})\right]\left(\frac{z}{z^{\star}}\right)^{\frac{n}{2}}\label{eq:Hzz}
\end{equation}
where $a$ is a constant determined by the boundary conditions.  Assuming a perfectly conducting boundary at $r=R$, and applying the condition of vanishing tangential electric field we can eliminate the unknown $a$ from (\ref{eq:Hzz})
\begin{align}
	\hat{\boldsymbol{\theta}}\cdot\boldsymbol{E}(z,z^{\star})&=-\frac{{\rm i}\eta_0}{k_0}\left[\frac{1}{\epsilon_{1}-\alpha}\frac{\partial H}{\partial z^{\star}}\sqrt{\frac{z^{\star}}{z}}+\frac{1}{\epsilon_1+\alpha}\frac{\partial H}{\partial z}\sqrt{\frac{z}{z^{\star}}}\right]_{\sqrt{zz^{\star}}=R}=0\nonumber\\[10pt]
    \to a&=-\frac{J_n'(kR)-\frac{n\alpha}{\epsilon_1 kR}J_n(kR)}{Y_n'(kR)-\frac{n\alpha}{\epsilon_1 kR}Y_n(kR)}\label{eq:general_boundary_condition}
\end{align}
\par
The dependence of the frequency $k_0$ on the material filling the cavity is then determined by the boundary conditions within the cavity in the three cases shown in figure~\ref{fig:topology_effect}, which are 
\begin{align}
	a&=0&\qquad&\text{Simply connected cavity}\nonumber\\
    a&=-\frac{J_n'(k\rho)-\frac{n\alpha}{\epsilon_1 k\rho}J_n(k\rho)}{Y_n'(k\rho)-\frac{n\alpha}{\epsilon_1 k\rho}Y_n(k\rho)}&\qquad&\text{PEC boundary at } r=\rho\nonumber\\
    a&=-\frac{J_n(k\rho)}{Y_n(k\rho)}&\qquad&\text{PMC boundary at } r=\rho.\label{eq:three_cases}
\end{align}
The lower portion of figure~\ref{fig:topology_effect} shows the dependence of $k_0$ on the value of $\alpha/\epsilon_1$ in these three cases.  These were computed numerically through looking for values of $k_0$ where (\ref{eq:general_boundary_condition}) equals (\ref{eq:three_cases}).
%
%

\end{document}